\documentclass[aps,prl,reprint,amssymb,showpacs,superscriptaddress,notitlepage,twocolumns]{revtex4-1}
\usepackage{bm}
\usepackage{graphicx,hyperref}
\usepackage{dcolumn}
\usepackage{braket}
\usepackage{natbib}
\usepackage{amsmath}
\usepackage{color}
\usepackage{siunitx}
\usepackage{mathtools,amssymb}
\usepackage{soul}

\definecolor{gray}{rgb}{0.7,0.7,0.7}
\definecolor{orange}{rgb}{1, 0.4, 0}
\definecolor{dgreen}{rgb}{0.0, 0.4, 0.0}
\definecolor{yblue}{rgb}{0.06, 0.3, 0.57}

\newcommand\ldsout{\bgroup\markoverwith{\textcolor{blue}{\rule[0.5ex]{2pt}{0.4pt}}}\ULon}

\begin{document}
\title{Spin-selective strong light-matter coupling in a 2D hole gas-microcavity system}
\author{D. G. Su\'arez-Forero}
\email{dsuarezf@umd.edu}
\affiliation{Joint Quantum Institute and Quantum Technology Center, NIST and University of Maryland, College Park, Maryland 20742, USA}
\author{D. W. Session}
\affiliation{Joint Quantum Institute and Quantum Technology Center, NIST and University of Maryland, College Park, Maryland 20742, USA}
\author{M. Jalali Mehrabad}
\affiliation{Joint Quantum Institute and Quantum Technology Center, NIST and University of Maryland, College Park, Maryland 20742, USA}
\author{P. Kn\"uppel}
\affiliation{Department of Physics, Cornell University, Ithaca, NY, USA}
\author{S. Faelt}
\affiliation{Solid State Physics Laboratory, ETH Zürich, CH-8093 Zürich, Switzerland}
\affiliation{Institute of Quantum Electronics, ETH Zürich, CH-8093, Zürich, Switzerland}
\author{W. Wegscheider}
\affiliation{Solid State Physics Laboratory, ETH Zürich, CH-8093 Zürich, Switzerland}
\author{M. Hafezi}
\affiliation{Joint Quantum Institute and Quantum Technology Center, NIST and University of Maryland, College Park, Maryland 20742, USA}
\affiliation{Pauli Center for Theoretical Studies, ETH Zürich, CH-8093, Zürich, Switzerland}

\begin{abstract}
The interplay between time-reversal symmetry breaking and strong light-matter coupling in 2D gases brings intriguing aspects to polariton physics. This combination can lead to polarization/spin selective light-matter interaction in the strong coupling regime. In this work, we report such a selective strong light-matter interaction by harnessing a 2D gas in the quantum Hall regime coupled to a microcavity. Specifically, we demonstrate circular-polarization dependence of the vacuum Rabi splitting, as a function of magnetic field and hole density. We provide a quantitative understanding of the phenomenon by modeling the coupling of optical transitions between Landau levels to the microcavity. This method introduces a control tool over the spin degree of freedom in polaritonic semiconductor systems, paving the way for new experimental possibilities in light-matter hybrids.
\end{abstract}

\maketitle
A gas of charged particles confined in two dimensions has been a cornerstone system to investigate numerous single-particle and many-body quantum mechanical effects. In these systems, the reduction in the dimensionality has enabled the demonstration of integer \cite{VonKlitzing1986,Klitzing1980} and fractional \cite{Stormer1992} quantum Hall effects, Mott-insulating phases \cite{Kawasugi2019,Li2021}, superconductivity \cite{Cao2018}, formation of skyrmions \cite{Schmeller1995,Shukla2000} and spin-ordered states \cite{Zuo2009} among other effects \cite{Bao2021}. Recently, the possibility of coupling such 2D gases to an optical mode, forming cavity exciton-polaritons, has opened up exciting new directions. For example, this capability has been exploited to optically study integer and fractional quantum Hall states \cite{Smolka2014,Ravets2018}, formation of skyrmions and manipulation of the magnetic Land\'e g factor \cite{Lupatini2020} and the enhancement of optical non-linearities, when an electron gas, in the fractional quantum Hall regime, is dressed with the cavity mode \cite{Knuppel2019}. The demonstration of the mentioned effects requires that the light-matter interaction strength in the microcavity exciton-polaritons exceeds cavity and matter dissipation --- a regime known as strong coupling (SC) \cite{carusotto2013quantum}.\\

Combining SC with broken time-reversal symmetry in chiral systems is expected to unlock new perspectives for the optical manipulation of correlated states of matter \cite{Whittaker2020,bloch2022strongly,Hubener}. The chirality can arise from either electronic or photonic degrees of freedom, and in the exciton-polariton systems, these two are intertwined. The present work illustrates a method to induce a strong spin selectivity in the light-matter coupling between an electronic transition and a cavity mode. To achieve this, we use a hole ($h^+$) gas in a semiconductor Quantum Well (QW), embedded in a microcavity. The system is subject to an external magnetic field ($B$) to be in the quantum Hall regime. The modification of the density of states induced by $B$ results in the selectivity of the light-matter coupling for each circularly polarized state of light. We explain this selectivity by combining an exciton-polariton model and the Landau level filling picture. \\

\begin{figure}
    \centering
    \includegraphics[width=0.99\columnwidth]{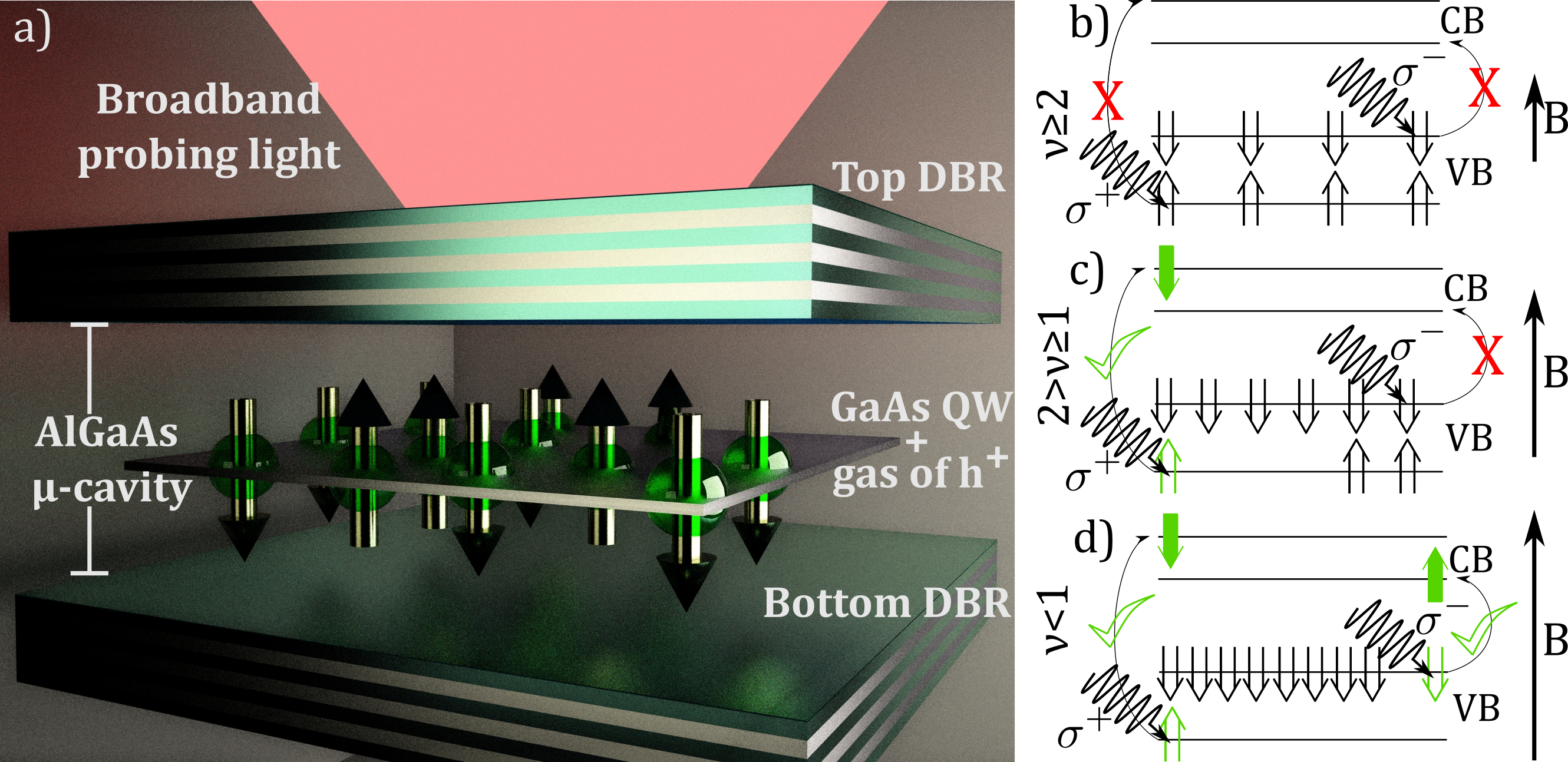}
    \caption{a) Diagram of the sample, including an illustration of the 2DHG embedded in an optical microcavity. b-d) mechanism to induce spin-selectivity for light-matter coupling benchmarked by three ranges of the filling factor $\nu$. VB and CB stand for Valence Band and Conduction Band, respectively. b) For $\nu\!\geq\!2$, none of the spin VBs has available states and therefore no electronic transition can take place. As a consequence, the SC regime will not be reached. c) As the magnetic field increases and enters the $1\!\leq\!\nu\!<\!2$ regime, the density of states is tuned such that the $\sigma^+$ sub-band becomes available. Unlike the former case, the electronic transition is not blocked anymore and this spin population reaches the SC regime. The $\sigma^-$ states are still Pauli-blocked, so no electronic transition is allowed for them. The result is the spin-selective light-matter coupling. d) As $B$ is further increased such that the system is tuned into the $\nu\!<\!1$ regime, the $\sigma^-$ band starts to have available states and can reach SC. The asymmetry in the Rabi frequency persists because the $\sigma^-$ band is still populated, in contrast to the $\sigma^+$ band that is fully available.}
    \label{diagram}
\end{figure}

Our physical system consists of a doped GaAs QW whose electronic states can be manipulated either through an applied gate voltage or a magnetic field. The sample is a p-doped grown QW, which constitutes a 2-dimensional $h^+$ gas (2DHG). To strongly enhance the interaction with an electromagnetic mode, the QW is embedded in a Distributed Bragg Reflector (DBR) semiconductor microcavity. Such a hybrid system provides a powerful tool to analyze the electron spin properties through optical experiments by exploiting the selection rules, which determine a spin selective transition depending on the circular polarization of light. The origin of this selectivity is in the crystalline structure and spin-orbit coupling properties of GaAs QWs, whose upper valence band is a heavy hole-band (with total angular momentum $J_z^h\!=\!\pm3/2$). A schematic representation of the 2D gas of semi-integer spin-charged particles confined to the QW and interacting with the electromagnetic field is displayed in Fig.~\ref{diagram}a. All presented data in the main text are obtained at 3.5K, where we do not expect any correlated electron physics to be present. A more detailed discussion of the selection rules and comparison to 40mK data can be found in the Supplementary Material.  The details of the sample composition and fabrication can be found in the Methods Section.\\

\begin{figure}[ht!]
    \centering
    \includegraphics[width=0.9\columnwidth]{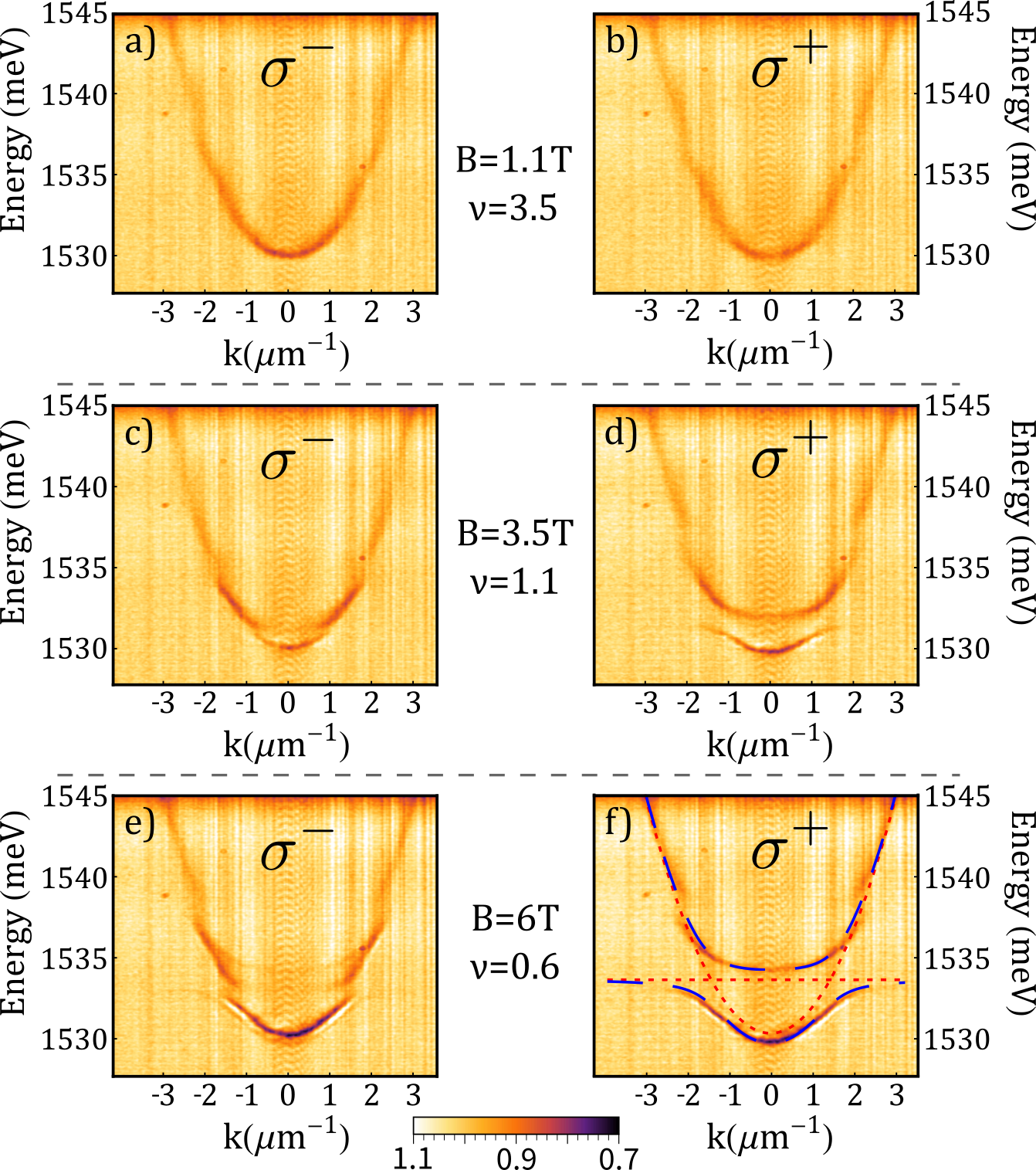}
    \caption{Energy dispersion of the system probed at values of $B$ corresponding to the three ranges of $\nu$ shown in Fig.~\ref{diagram}b-d. Panels a, c, and e correspond to $\sigma^-$ polarization, and panels b, d, and f to $\sigma^+$. a) and b): when $\nu\geq2$ both spin sub-bands are occupied and no electronic transition can take place in the system. Only the bare cavity mode is visible. c) and d): in the regime $2>\nu\geq1$ only the $\sigma^+$ spin sub-band has available states, and hence the spin selectivity manifests. Only the dispersion probed with $\sigma^+$ polarization shows polaritonic bands, while $\sigma^-$ remains decoupled. e) and f): upon further increasing $B$, also the $\sigma^-$ population can reach the SC, but the remaining $h^+$ population in this sub-band makes the light-matter coupling asymmetric, making the Rabi splitting larger for $\sigma^+$. Panel f includes the fitting obtained from a theoretical model (Eq.~\ref{hamiltonian}). Cavity and exciton modes are depicted in red dotted lines, while upper and lower polariton branches are depicted in dashed blue lines.} 
    \label{asymmetry}
\end{figure}

An external magnetic field acting on the 2D gas induces Landau levels that are non-degenerate for the spin eigenstates. This spin energy splitting depends on the magnitude of $B$, and in particular, for a $h^+$ doped system, it has a strong dependence on the particle density due to the spin-orbit interaction  \cite{Winkler2003,Ma2022}. At the same time, the $h^+$ density and the magnetic field determine the Fermi energy and hence the existence of an optically active transition of the electrons in the valence band \cite{Efimkin2018,Rana2020}. We exploit this particular interplay to obtain a spin-selective light-matter interaction. Specifically, for a fixed
density of particles, we tune $B$ to consider three relevant situations with fundamentally different light-matter coupling properties. The quantity that determines the three cases is the quantum Hall filling factor $\nu\!=\!\rho/n_B$, defined as the ratio between the $h^+$ density $\rho$ and the density of magnetic states $n_B\!=\!eB/h$ ($e$ is the electron charge and $h$ is Planck's constant). The three cases are depicted in Fig.~\ref{diagram}(b-d). The first scenario (b) takes place when $\nu\!>\!2$. In this condition, there are no available states in any of the two spin sub-bands, inhibiting the possibility of an electronic transition from the lowest Landau level and therefore keeping the system in the so-called weak coupling regime. The next situation takes place when $1\!\leq\!\nu\!<\!2$ (c). In this case, one of the sub-bands ($\sigma^-$) is still full, but the opposite spin state ($\sigma^+$) can instead host optically-generated electron-hole pairs, leading to a polarization-selective SC. Finally, when $\nu\!<\!1$, as shown in Fig.~\ref{diagram}(d), both spin transitions are open, but due to the imbalance in the $h^+$ population, the system still has a strong asymmetry in the light-matter coupling. \\

To investigate the proposed mechanism, we perform angular and spectrally-resolved reflectivity measurements to reconstruct the energy dispersion in a confocal setup. Such a setup is capable of mapping the far-field for any polarization state, using a broadband light source for excitation. A complete description of the optical setup can be found in the Methods section. $\rho$ is manipulated through a gate voltage and it is set to $\sim\!9.4\!\cdot\!10^{10} cm^{-2}$; a density high enough to guarantee that at low $B$, both spin valence sub-bands ($\sigma^+$ and $\sigma^-$) are occupied, but low enough to allow the desired effect to take place at a few Teslas, i.e., the higher the density, the higher the magnetic field at which the conditions $\nu\!=\!2$ and $\nu\!=\!1$ take place. We sweep $B$ to tune the system over a range of filling factors that covers the three regimes described in Fig.~\ref{diagram}(b-d).~Figure \ref{asymmetry} shows the measured dispersion of the exciton-polariton system for both circular polarizations and values of $B$ in the three ranges of $\nu$. For $\nu\!\geq\!2$, as shown in panels a and b, and in agreement with the above description, only the parabolic dispersion from a bare cavity mode can be observed. This indicates that none of the spin sub-bands have available states and hence no optical electronic transition can take place. Next, by increasing $B$, we reach the $2\!>\!\nu\!\geq\! 1$ regime, and then the asymmetry in SC is manifested. While the dispersion is still parabolic for $\sigma^-$ (panel c), the $\sigma^+$ channel has clearly reached the SC regime, as observed from the formation of lower and upper polariton branches in the energy dispersion (panel d). By further increasing $B$, we reach the regime $\nu\!<\!1$ (panels e and f), where both spin sub-bands can now reach SC, but the asymmetry in the $h^+$ distribution causes the population of $\sigma^-$ to couple less efficiently to the cavity mode than the population in the $\sigma^+$ sub-band. As a consequence, the vacuum Rabi splitting is different for each state of circular polarization.\\

A coupled oscillator model is employed for the quantitative analysis of our observation, characterized by the polaritonic Hamiltonian: 
\begin{equation}{
\hat{H}=\sum_k(\omega_{\rm{c}}+\frac{k^2}{2m})\hat{a}_k^{\dagger}\hat{a}_k+\omega_{\rm X}\hat{b}^{\dagger}\hat{b}+\Omega\sum_k\hat{b}^{\dagger}\hat{a}_k+\hat{b}\hat{a}_k^{\dagger},}\label{hamiltonian}\end{equation}
for each polarization, where $\omega_{\rm c}$ is the cavity mode's energy, $k$ its momentum, $m$ its effective mass, and $\omega_{\rm X}(B)$ and $\Omega(B)$ are the energy of the electronic transition and the Rabi splitting, respectively (the latter two quantities depend on $B$). The operators $\hat{a}_k$ and $\hat{b}$ correspond to the annihilation operators of the cavity mode $k$, and electronic transition, respectively \cite{Kavokin2017}. An example of the obtained polaritonic eigenfunction after fitting the data is included in Fig.~\ref{asymmetry}f. The cavity and exciton modes are depicted (dotted red lines) together with the upper and lower polariton branches (dashed blue lines) that match the experimental dispersion.\\

\begin{figure}
    \centering
    \includegraphics[width=0.99\columnwidth]{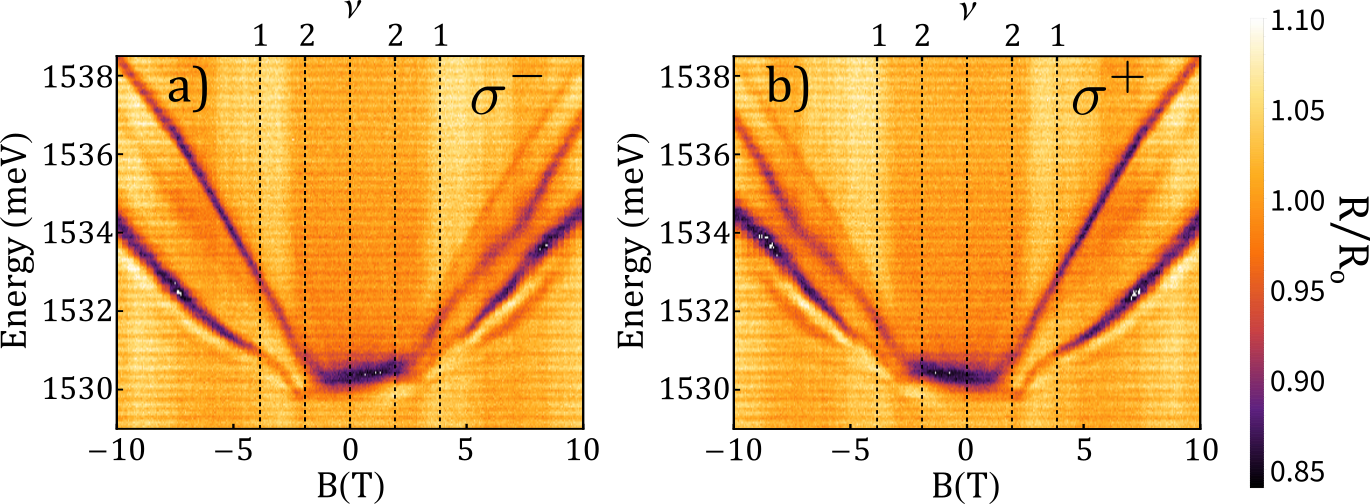}
    \caption{Magnetic field dependence of the system resonance at momentum at which the bare cavity and exciton energies cross. Due to both Zeeman splitting and diamagnetic shift, this crossing occurs at different momentum for each $B$. The values of $B$ for filling factors 1 and 2 that delimit the three regimes of the system are indicated by black dashed lines for positive and negative values of $B$. The data is displayed for both spin states (a) $\sigma^-$, (b) $\sigma^+$.}
    \label{Energy-B}
\end{figure}

The trend of spin-selective light-matter coupling as a function of the filling factor is illustrated in Fig.~\ref{Energy-B}. To obtain it, we extract a vertical section of the energy dispersion corresponding to the momentum at which the bare cavity and electronic transition cross (deduced from the theoretical model). From Fig.~\ref{Energy-B} one can observe how the filling factor $\nu\!=\!2$ benchmarks the establishment of SC for $\sigma^+$, which is observed as the emergence of two resonances instead of one, while $\sigma^-$ remains decoupled. Only for filling factors above $\nu\!=\!1$ the two polariton resonances can be observed for this spin state, although, as previously mentioned, there is a persistent difference in the value of $\Omega$. Here it is worth noting that as a consequence of imperfections in the polarization filters, a feeble trace of the opposite spin reflection contrast can be detected on each panel of Fig.~\ref{Energy-B}. The evolution of the full energy dispersion with a magnetic field can be found in Supplementary Video 1.\\

\begin{figure}
    \centering
    \includegraphics[width=\columnwidth]{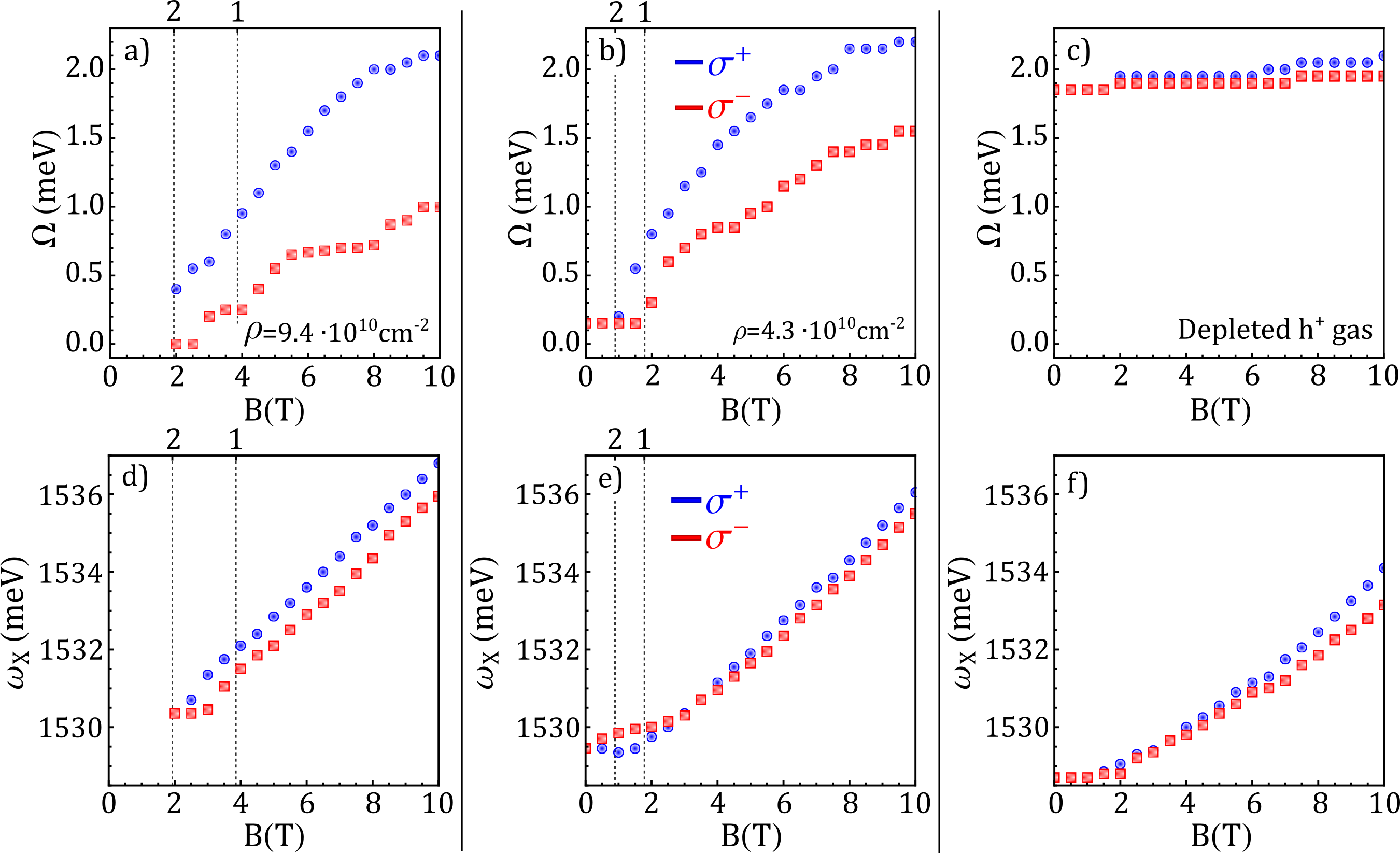}
    \caption{a-c) Rabi splitting obtained from the fitting to a polariton model (equation \ref{hamiltonian}) for selected magnetic fields and for three different $h^+$ densities. d-f) Energy of the electronic transition obtained from the theoretical model for the same selected magnetic fields and densities. a) and d) $\rho=9.4\cdot 10^{10}$ $cm^{-2}$. b) and e) $\rho=4.3\cdot 10^{10}$ $cm^{-2}$. c) and f) fully depleted $h^+$ gas. As expected, the light-matter coupling selectivity is not observed in this last situation.}
\label{exc-ome}
\end{figure}

To obtain a quantitative description of the spin-selective SC effect, we extract the values of $\omega_{\rm X}$ and $\Omega$ from the described theoretical model for each magnitude of $B$ and for each spin state. The results are displayed in Fig.~\ref{exc-ome}. From the Rabi splitting value (upper panels), one can observe how the integer filling factors (marked on the top axis) coincide with the magnetic fields at which the populations $\sigma^+$ and $\sigma^-$ reach the SC: $\nu\!=\!2$ and $\nu\!=\!1$, respectively. As previously mentioned, for $\nu\!<\!1$ the asymmetry in $\Omega$ persists due to the $h^+$ population in the $\sigma^-$ sub-band; we expect that for even higher magnetic fields, the difference will be asymptotically eliminated as $\nu\!\xrightarrow{}\!0$. Notice that for $\nu\!<\!1$, $\Omega$ further increases for the $\sigma^+$ polariton resonances. This is due to the electron and hole confinement induced by the magnetic field that increases the oscillator strength and hence $\Omega$ \cite{Cong2018}. Finally, for a fully depleted $h^+$ gas (panel c) the selectivity is completely removed, and both spin sub-bands are in SC with the cavity mode, in agreement with the expected behavior of a polaritonic system. The supplementary videos show the fitting of the energy dispersion upon a variable magnetic field, for the densities studied in this work. \\

The theoretical model (equation \ref{hamiltonian}), in close agreement with the experimental observation, provides the value of the electronic transition energy (Fig.~\ref{exc-ome}d-e). However, it is important to mention that a complete comprehension of the nature of this optically induced transition requires further experimental and theoretical investigation. Although the existence of this transition depends on the filling factor, its microscopic properties are determined by the interplay between: 1.~electron-hole Coulomb interaction, 2.~interaction with the 2DHG, and 3.~interaction with the external magnetic field. Since none of these interactions dominates over the other two, a complete description of this polaritonic quasi-particle is not straightforward. Indeed, recent theoretical efforts are directed to investigate the nature of the light-matter coupling under these circumstances \cite{Laird2022}. It is also important to highlight that although the nature of this quasi-particle is still an object of active research, the demonstration of the selective coupling to the cavity mode does not depend on it. Hence, the results presented in this work, remain valid regardless of the nature of the matter excitation.\\

 An asymmetry in the light-matter coupling has been reported for similar systems at lower charge densities when the 2DHG is in a spin-ordered state in the frame of integer or fractional quantum Hall effects \cite{Ravets2018,Lupatini2020,Chervy2020}. In those cases, the asymmetry takes place for the specific values of $B$ at which correlated states of matter dominate. 
This is not the case for the studied system in the present work. In our case, any trace of correlated matter states at integer filling factors is ruled out by increasing the temperature ($T$) of the system to 3.5K (in contrast to  Refs.~\cite{Ravets2018,Lupatini2020} where $T$ is kept at $\sim$100$mK$). The fact that the presented results are collected at a relatively high $T$ (3.5K) indicates that our observation is not related to many-body electronic correlations, highlighting the robustness of this mechanism. The Supplementary Material includes a comparison between the system's response at 40mK and 3.5K as well as the method for calibrating the $h^+$ density that requires these ultra-low $T$.\\

In summary, we introduced an experimental method for spin-selective light-matter coupling in a microcavity-quantum well system resulting from the manipulation of the spin valence sub-bands through a modification of the $h^+$ density and the application of a magnetic field. This result opens perspectives on the control of non-linear effects in these systems, specifically, the possibility to directly manipulate the value of $\chi^{(3)}$ using the polarization of the light: a system in SC has a high non-linearity inherited from the excitonic component, in contrast with the purely photonic case, where the exciton-exciton interaction does not take place, and hence, the non-linearity reduces dramatically. Additionally, the possibility of application of this method is not limited to III-V semiconductor devices. One potential alternative platform is TMD materials where spin and valley degrees of freedom are locked. This method, combined with the possibility of embedding a monoatomic semiconductor in a microcavity \cite{Liu2015} opens interesting perspectives for a valley selective light-matter SC. A possibility that may be more accessible after recent experiments that demonstrated giant Zeeman splitting in a monolayer TMD \cite{Lyons2022}, enhanced by exchange interactions in the presence of free carriers \cite{Back2017}. This system might also enable technological implementations such as variable circular polarization filters with ultra-narrow bandwidth, and from a more fundamental point of view, it is a suitable platform to optically study the spin physics of 2D gases of charged fermions, thanks to the degree of control on the light-matter interaction that can be achieved by modifying the $h^+$ density.

\subsection{Methods}
\noindent \textbf{Sample fabrication and preparation}

Our sample consists of a $\rm{Al}_{0.2}\rm{Ga}_{0.8}\rm{As}$ $\lambda$ cavity embedded between two $\rm{AlAs}$/$\rm{Al}_{0.2}\rm{Ga}_{0.8}As$ DBRs (24 bottom pairs and 19 top pairs) with a p-doping (Carbon) layer and a 16nm GaAs QW, grown by molecular beam epitaxy. To control the $h^+$ density, a 60 nm Si-doped n-type GaAs layer is grown between the GaAs substrate and the bottom DBR. This acts as a gate that can control the density of the 2DHG. Contacts were fabricated using In for the n-type GaAs gate and $\rm{In}_{0.96}\rm{Zn}_{0.04}$ for the 2DHG. The device's $h^+$ density can then be controlled by applying a gate voltage.\\

\noindent\textbf{Optical measurements}

The sample is placed in a dilution refrigerator with the capability of reaching temperatures down to $~$40 mK. Although all the measurements presented in the manuscript are taken at a temperature of 3.5K, we perform measurements at a base temperature of $~$40mK to calibrate the $h^+$ density, using the optical signature of correlated states at integer filling factors. The comparison of the data taken at base temperature and 3.5K as well as the details of the density calibration procedure can be found in the Supplementary Material.\\ 
For the optical measurements, we use a confocal setup in reflectivity configuration with numerical aperture $\rm{NA}=0.5$. A tungsten lamp is employed as a broadband light source, but consistent results were obtained using a supercontinuum white light source. Polarization-resolved spectra were collected by placing a linear polarizer followed by a quarter wave plate in the excitation path.

\subsection{Acknowledgements}
The authors thank Atac Imamoglu, Sylvain Ravets for valuable discussions and enriching feedback on the manuscript. This work was supported by AFOSR FA9550-20-1-0223, FA9550-19-1-0399, and ONR N00014-20-1-2325, NSF IMOD DMR-2019444, ARL W911NF1920181, Minta Martin and Simons Foundation.

\subsection{Data availability}
All data are available from the corresponding authors upon reasonable request.

\bibliography{Biblio.bib}

\newpage

\setcounter{equation}{0}
\setcounter{figure}{0}
\setcounter{table}{0}
\setcounter{page}{1}
\makeatletter
\renewcommand{\theequation}{S\arabic{equation}}
\renewcommand{\thefigure}{S\arabic{figure}}
\pagenumbering{roman}

\noindent
\textbf{\Large Supporting Information}\\

\section{Contents}
\begin{enumerate}\bf{
    \item Spin selection rules for optical transitions
    \item Hole density calibration
    \item Comparison of the system response at T=40mK vs T=3.5K
    \item Details on the theoretical formalism for data fitting}
\end{enumerate}

\subsection{1. Spin selection rules for optical transitions}

A fundamental condition, necessary for the spin-selective strong coupling reported in this work, is the existence of spin-selection rules that determine the optical transitions that can take place in the system. In this section, we will discuss the nature and origin of these selection rules, mostly based on Ref.~\cite{Kavokin2017}. They originate from the dipolar nature of the light-matter interaction which in turn arises from the spin conservation in photoabsorption events. Due to the strong spin-orbit coupling, the valence band (VB) energy is non-degenerate in quantum wells with a Zincblende crystalline structure. The VBs are composed of a band with a total angular momentum $J_{z}^{h}=\pm3/2$ usually referred to as the ``heavy-hole'' (HH) band and one with $J_{z}^{h}=\pm1/2$, referred to as ``light-hole'' (LH) band. It is the relative orientation of the spin and the orbital angular momentum which determines the nature of the hole band. In typical GaAs structures, the HH band is closer to the conduction band (CB) than the LH band, hence, in this analysis, we focus on the lowest energy optical transitions, i.e. those between the HH band and the CB.\\

The optical transition has a total angular momentum projection corresponding to the projection of the addition of angular momentum of electron ($J_{z}^{e}=\pm1/2$) and HH ($J_{z}^{h}=\pm3/2$). This transition can then have projections of the total angular momentum $J_{z}^{T}=\pm1,\pm2$. The angular momentum conservation implies that only those transitions matching the photon's angular momentum $\pm1$ can take place, hence, only the configurations in which the projections of electron and hole are anti-parallel are optically allowed. Typically, for a neutral QW system, the exciton corresponding to this projection is called ``bright'' exciton, in contrast to the $J_{z}^{T}=\pm2$ case, referred to as ``dark'' exciton.\\

\begin{figure}
    \centering
    \includegraphics[width=0.9\columnwidth]{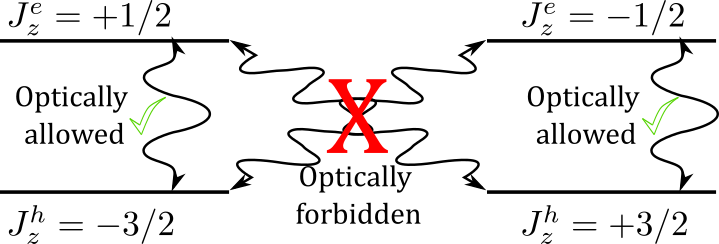}
    \caption{Optical selection rules for a GaAs quantum well for transitions between the HH band and the CB.}
    \label{rules}
\end{figure}
These selection rules determine the existence of spin selectivity in the strong coupling. Figure \ref{rules} shows a diagram of the optical selection rules. It shows how the transition from the spin up (down) VB to the spin up (down) CB is optically forbidden, which justifies why these transitions are not indicated in Fig.~1b-d of the main document.

\subsection{2. Hole density calibration}
\begin{figure*}
    \centering
    \includegraphics[width=1.9\columnwidth]{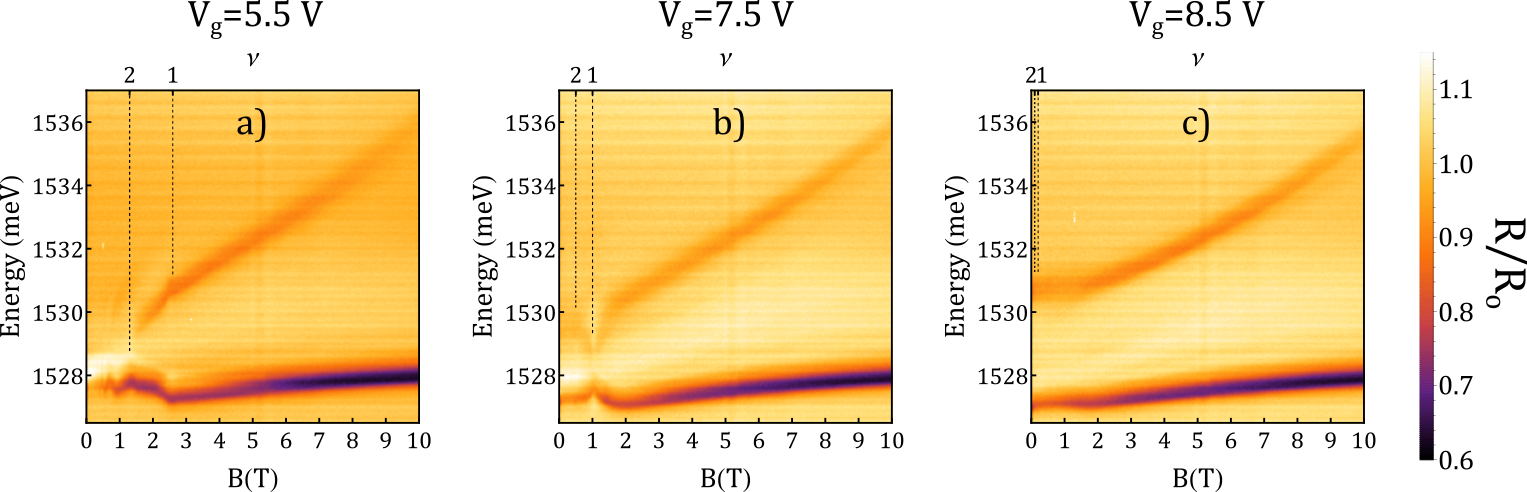}
    \caption{Differential reflectivity of the system at 40 mK. The displayed data corresponds to low numerical apertures ($k\sim0$). The gate voltage sets the hole density and a sweep over the magnetic field allows us to identify the correlated states of matter associated with integer filling fractions. Due to the finite hole mobility of the sample, the effect is not visible for every density. In the figure, one can observe that for high densities, the modification of the absorption spectrum corresponding to $\nu=1$ and $\nu=2$ can be observed (panel a). For medium densities, only $\nu=1$ shows the feature (panel b) and for low densities, none of the correlated states is observable (panel c). The top axis indicates the magnetic fields corresponding to integer filling fractions obtained from an interpolation of the results.}
    \label{opt_sign}
\end{figure*}

At a temperature of 40 mK, the hole gas is capable of sustaining correlated states of matter at integer filling fractions. As mentioned in the main text, this result has been previously reported \cite{Lupatini2020}. This gives a very accurate method for calibrating the hole density: since the magnetic field strength and the gate voltage are known quantities, one can infer the hole density from the optical signatures of the integer filling factor states $\nu=1$ or $\nu=2$. According to quantum Hall physics, these correlated states are established when $\rho h/eB \in \mathbb{Z}$. Depending on $\rho$, one can observe both correlated states, one of them, or none. This limitation is imposed by the finite hole mobility in the sample. Figure \ref{opt_sign} shows the reflectivity spectrum of the system at low numerical apertures i.e.~close to zero in-plane momentum, as a function of the magnetic field. After systematically collecting data for different gate voltages, and recording the magnetic fields at which the Rabi splitting is modified, one can obtain an accurate linear model that gives the hole density of the system as a function of the gate voltage. 

\begin{figure}
    \centering
    \includegraphics[width=0.95\columnwidth]{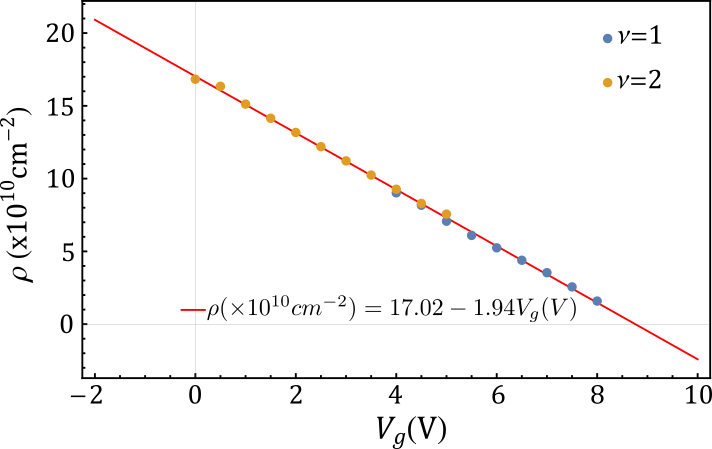}
    \caption{Voltage to hole density calibration. From the low momentum reflectivity spectrum, we extract the values of the magnetic fields for which the correlated states of matter take place for each voltage. In some cases both ($\nu=1$ and $\nu=2$) states are observable, in some cases only one of them is, and for very high and very low densities, none of the correlated states are observable. The red line shows the fitting line for the data set, which we use as a reference to obtain the hole density at 3.5 K. The range of the horizontal axis corresponds to the range at which we collected data.}
    \label{density}
\end{figure}

The density is not perfectly uniform across the sample; we performed this calibration in different spots obtaining different results. For this reason, it is important to perform a calibration procedure every time the position of the excitation spot on the sample changes. Figure \ref{density} shows the deduced hole density for each given gate voltage at a given spot of the sample. The procedure consists in registering the magnetic field at which the Rabi splitting is modulated for a given gate voltage. With this information, we infer the hole density according to the formula  $\rho=eB\nu/h$. $\nu$ taking the values $1$ or $2$. Notice from the figure that for some gate voltages both integer filling fractions are displayed, because in those cases the optical signatures of both states are observable. This corresponds, for example, to the case displayed in Fig.~\ref{opt_sign}a. For other gate voltages, only one of the correlated states can be detected as shown, for example, in Fig~\ref{opt_sign}b. However, for very high or very low densities, none of the correlated states can be observed (Fig.~\ref{opt_sign}c). After recording the calculated density for each gate voltage, we use the best-fitting linear function as a reference for the system's density once the temperature is increased from 40 mK to 3.5 K.\\

\begin{figure}
    \centering
    \includegraphics[width=\columnwidth]{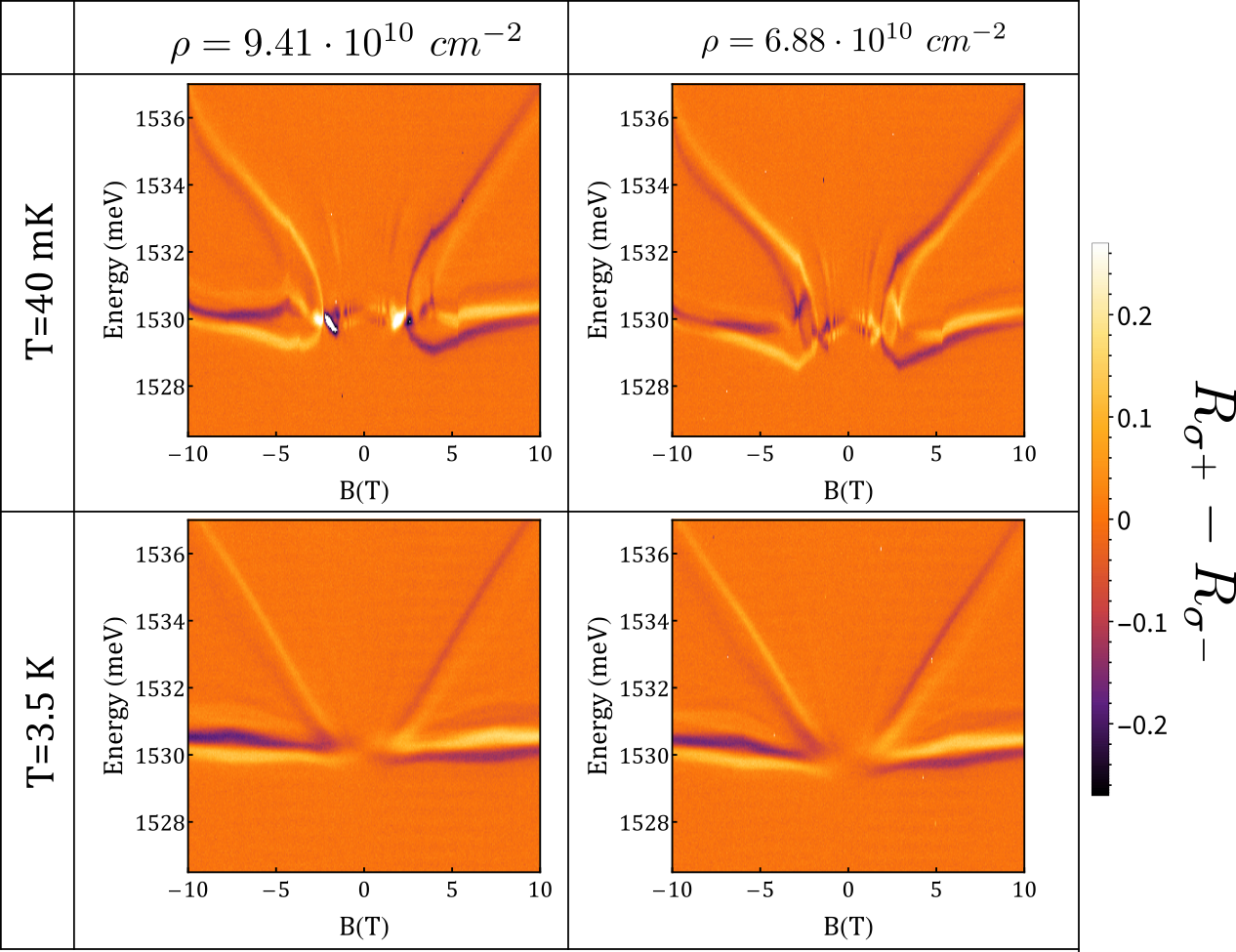}
    \caption{Difference of the reflectivity spectra for the two circular states of polarization. The data is presented for two densities ($9.41\cdot10^{10}$$\rm{cm}^{-2}$ and $6.88\cdot10^{10}$$\rm{cm}^{-2}$) and for two temperatures (40 mK and 3.5 K). The modulation of the resonance energy peak associated with correlated states of matter is visible only for the data collected at the dilution unit base temperature (40 mK).}
    \label{4Kvs40mK}
\end{figure}
\subsection{3. Comparison of the system response at T=40mK vs T=3.5K}
In this section we discuss the fundamentally different response of the physical system when probed at 40 mK or at 3.5 K. This is important because the previously reported modulation of the Rabi splitting \cite{Lupatini2020,Smolka2014,Ravets2018} relies on correlated states of matter that can only take place at these ultra-cold temperatures, while the effect we exploit to induce the spin-selective strong coupling does not rely on this physics. As mentioned in the main text, the optical features from the correlated states at integer filling fractions are completely washed out when the temperature is 3.5 K. A magnetic field dependence was collected in identical conditions at 40 mK and 3.5 K for two densities: ($9.4\cdot10^{10} cm^{-2}$ and $6.88\cdot10^{10} cm^{-2}$). The results, displayed in Fig.~\ref{4Kvs40mK} respond to the difference of the reflectivity spectra for $\sigma^+$ and $\sigma^-$ polarization collected with low numerical aperture. The subtraction of the reflectivity spectra allows us to show both of the spin (polarization) states in a single frame. In contraposition with the data presented in Fig.~3 of the main text, the in-plane momentum is constant over this set of data ($k\sim0$).\\

The figure demonstrates the fundamental difference in the sample response when changing the temperature. As previously mentioned, upon a fixed hole density, the magnetic field tunes the system's filling factor. For the data collected at 40 mK (Fig.~\ref{4Kvs40mK} upper panels), at specific values of the magnetic field, the Rabi splitting is strongly modified. This stems from the interaction-driven modulation of the hole spin polarization and changes in the available states for optical transitions as the filling factor is varied. In the lower panel of Fig.~\ref{4Kvs40mK}, we compare the system response at the same two densities, but at a higher temperature of 3.5 K. From the results, it is evident that the thermal noise overcomes the correlations present in the system, washing out any trace of a modulation of the system's Rabi splitting. As pointed out in the main text, the spin selectivity reported in this work is robust against thermal effects, hence, it is observable even at 3.5K.

\subsection{4. Details on the theoretical formalism for data fitting}
The coupled oscillators formalism employed for the theoretical modelization of the polaritonic system is described by the Hamiltonian\begingroup\makeatletter\def\f@size{8}\check@mathfonts
\begin{equation}{\small
\hat{H}=\sum_k(\omega_{\rm{c}}+\frac{k^2}{2m})\hat{a}_k^{\dagger}\hat{a}_k+\omega_{\rm X}(B)\hat{b}^{\dagger}\hat{b}+\Omega(B)\sum_k(\hat{b}^{\dagger}\hat{a}_k+\hat{b}\hat{a}_k^{\dagger})}\label{hamiltonian}\end{equation}\endgroup
As specified in the main text, $\omega_{\rm c}$ is the cavity mode's energy, $k$ its momentum, $m$ its effective mass, and $\omega_{\rm X}(B)$ and $\Omega(B)$ are the energy of the electronic transition and the Rabi splitting, respectively \cite{Kavokin2017}. Once collected, we fit the data by using $\omega_X$ and $\Omega$ as fitting parameters for each magnetic field. The data is then fitted to the eigenfunctions of the hamiltonian \ref{hamiltonian}, which are given by: \begingroup\makeatletter\def\f@size{10}\check@mathfonts
\begin{equation}{E_{U/L}=\frac{1}{2}\left(\omega_X+\omega_{\rm{c}}+\frac{k^2}{2m}\right)\pm\frac{1}{2}\sqrt{\Delta_k^2+4\Omega^2}
}\label{eigenstates}\end{equation}\endgroup With $\Delta_k=\omega_{\rm{c}}+\frac{k^2}{2m}-\omega_X$. The cavity's energy and effective mass is accurately extracted from the case $\Omega=0$ since in this condition, only the bare cavity mode is visible. For the mass, we obtain the value $\rm{m}=2.4\cdot10^{-5}\rm{m_e}$, with $\rm{m_e}$ the free electron mass. This value is kept constant over the fitting procedure of the full data set and it is consistent with typical values for these devices. The cavity energy accounts only for minor changes due to the response of the optical properties of the bare electromagnetic mode to the high magnetic and electric fields. In the absence of any field, the cavity mode at 0 in-plane momentum has energy $\omega_{\rm{C}}\!=\!1530$ meV, and for the full set of data (0V to 10V and -10T to 10T), it never changes by more than $0.5$ meV.\\

This model accurately reproduces the experimental data as it can be observed in the Supplementary videos, which show the evolution of the far-field reflectivity with an increasing magnetic field for different hole densities. The main document also displays one fitted set of data in Fig.~2f.\\

\bibliography{Biblio}

\end{document}